%% file: main.tex
\documentclass[conference,A4]{IEEEtran}

\IEEEoverridecommandlockouts

\usepackage[utf8]{inputenc}
\usepackage[T1]{fontenc}
\usepackage{url}
\usepackage{ifthen}
\usepackage{cite}
\usepackage[cmex10]{amsmath} %
\usepackage{bbm}
\usepackage{amssymb, amsfonts}
\usepackage{nicefrac}
\usepackage{aligned-overset}
\usepackage{graphicx}         %
\usepackage{subcaption}
\usepackage[nolist]{acronym}
\usepackage[normalem]{ulem}
\usepackage{pifont}
\usepackage{booktabs}
\usepackage{tabularx}
\usepackage{multirow}
\usepackage{array}

\usepackage{tikz}
\usetikzlibrary{arrows,calc,shapes,graphs,graphs.standard,quotes,arrows.meta,decorations.markings,positioning,fit,decorations.text,decorations.pathreplacing, dsp}

\usepackage{pgfplots}
\pgfplotsset{compat=1.17}
\pgfdeclarelayer{background}
\pgfdeclarelayer{foreground}
\pgfsetlayers{background,main,foreground}
\usepgfplotslibrary{groupplots}

\usepackage{amsthm}

\newtheorem{theorem}{Theorem}%
\newtheorem{example}{Example}%

\input{acronyms}
\input{colours}
\input{macros}

\begin{document}

\title{Serial Polar Automorphism Ensemble Decoders\\for Physical Unclonable Functions}

\author{\IEEEauthorblockN{Marvin Rübenacke\textsuperscript{\dag,*},
		Sebastian Cammerer\textsuperscript{\ddag},
		Michael Sullivan\textsuperscript{\ddag}, and
		Alexander Keller\textsuperscript{\ddag}}
\IEEEauthorblockA{
   \textsuperscript{\dag}University of Stuttgart,
   \textsuperscript{\ddag}NVIDIA, contact: scammerer@nvidia.com}\\
\thanks{\textsuperscript{*}Work done during an internship at NVIDIA.}
}

\maketitle

\begin{abstract}
\Acp{PUF} involve challenging practical applications of \acp{ECC}, requiring extremely low failure rates on the order of 10$^{-6}$ and below despite raw input bit error rates as high as 22\%.
These requirements call for an efficient ultra-low rate code design.
In this work, we propose a novel coding scheme tailored for \acp{PUF} based on Polar codes and a low-complexity version of \ac{AED}.
Notably, our serial \ac{AED} scheme reuses a single \ac{SC} decoder across multiple decoding attempts. By introducing cascaded and recursive interleavers, we efficiently scale the number of \ac{AED} candidates without requiring expensive large multiplexers. An aggressive quantization strategy of only 3 bits per message further reduces the area requirements of the underlying \ac{SC} decoder.
The resulting coding scheme achieves the same block error rate of 10$^{-6}$ as our baseline based on  \ac{BCH}  codes while requiring 1.75$\times$ fewer codeword bits to encode the same $K=312$ payload bits. This reduction translates directly into 1.75$\times$ less helper data storage and, consequently, a smaller overall chip area.

\end{abstract}

\section{Introduction}\label{sec:intro}
\acresetall

Physical Unclonable Functions (PUFs)\acused{PUF} generate a unique and unpredictable hardware fingerprint across semiconductor devices based on random variations of the manufacturing process that cannot be cloned \cite{suh2007physical,gunlu2019codeconstruction, bloch2021overview}. Such perfectly secret information provides protection against copying, cloning, and readout by attackers---even when using intrusive or destructive techniques.
One physical realization of a \ac{PUF} involves reading bits from uninitialized static random-access memory (SRAM).
While random across devices, an individual cell tends to output the same binary value when queried multiple times.
Carefully combining the noisy
\ac{PUF} responses with a strong \ac{ECC} scheme can guarantee reliable data extraction without leaking private information through helper data \cite{delvaux2015helperdataalgos, gunlu2019codeconstruction}. As the reliability even depends on hardware parameters such as temperature and cell aging, heavily overdesigned \ac{ECC} schemes are necessary. %

Traditional approaches in academia mostly focus on concatenated codes that combine low-complexity repetition with modest-complexity \ac{BCH} codes \cite{delvaux2015helperdataalgos, maes2016biasedpufs}. Such concatenated codes work well in moderate error-rate regimes but they do not scale gracefully to very high raw input error rates. As an alternative, Polar codes \cite{arikan2009} have been proposed for \acp{PUF} \cite{chen2017robustpolarpuf}, promising higher code rates, though previous work mostly focused on \ac{SC} decoding since the hardware complexity of \ac{SCL} decoding appears prohibitive for practical \ac{PUF} applications.

In this work, we propose a serial variant of \acf{AED}. \ac{AED} was originally developed in the context of wireless communications \cite{geiselhart2021aedrm, geiselhart2021polaraut,pillet2021polarcodesaed,kestel2023URLLC}, and it can achieve near-\ac{ML} decoding performance by applying multiple decoding attempts using permuted versions of the noisy codeword. However, scaling the number of \ac{AED} candidates is potentially expensive, as it either requires many parallel \ac{SC} decoders or large multiplexers for the interleavers/de-interleavers \cite{ren2024highthroughputbpl}.

For efficient hardware re-use, the interleavers must be carefully optimized to avoid large multiplexers.
In \cite{li2025routes}, the authors optimize the permutations to reduce implementation overhead, but do not consider their impact on decoding performance.
We propose cascaded interleavers based on the fact that permutation automorphisms form a group, and thus a cascade of interleavers will produce another valid permutation of the original codeword. Additionally, even a single interleaver can recursively generate multiple permutations, which further reduces the area footprint while not compromising the decoding performance.

The main contributions of this work are a
\begin{itemize}
	\item novel serial Polar \ac{AED} reusing a single \ac{SC} decoder,
	\item cascaded and recursive interleavers for \ac{AED} that efficiently scale the number of candidates,
	\item binary input \ac{AED} with aggressive 3-bit quantization, and
	\item \ac{VLSI} synthesis results demonstrating area efficiency.
\end{itemize}

Throughout this work, vectors and matrices are denoted by boldface lowercase $\xv$ and uppercase $\Am$ letters, respectively. $S_N$ denotes the symmetric group of degree $N$. For a permutation $\pi \in S_N$, $j = \pi(i)$ maps the indices $i\in \{0,1,
\dots, N-1\}$ to $j\in \{0,1,\dots, N-1\}$, and we write $\xv' = \pi(\xv)$ to
denote the reordering of a vector $\xv \in \Sc_N$ according to $ x'_{\pi(i)} =
x_i$.

\section{System Model and Background}\label{sec:prelim}

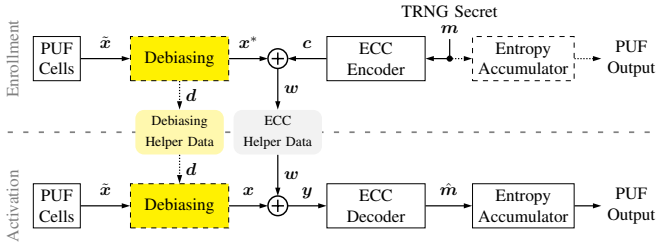
\begin{figure}
	\centering
	\resizebox{\linewidth}{!}{\input{fig/fuzzy_commitment}}
	\caption{ Overview of the \ac{PUF} key extraction pipeline using the fuzzy commitment scheme. The debiasing (yellow) is an optional component.}
	\label{fig:puf_overview}
\end{figure}

In the following, we introduce the serial version of AED in the context of \acp{PUF} which motivates the need for ultra-low rate codes and high area efficiency. Note that the proposed scheme can also be applied to any other communication or storage system. Fig.~\ref{fig:puf_overview} shows an overview of the system.

\subsection{Physical Unclonable Functions}

A \ac{PUF} is modeled by
$	\xv = \xv^* \oplus \ev $
and consists of $N$ binary \ac{PUF} cells $x_i \in \FF_2$.
The ground truth vector $\xv^*$ is called the \emph{golden PUF response} that is only observable through a noisy channel.
This channel is assumed to be a \ac{BSC} with bit-flip probability $\epsilon$ and a noise vector $\ev$ with $e_i \sim \operatorname{Ber}(\epsilon)$.

\subsection{Helper Data Algorithms}
To cope with the errors, an \ac{ECC} is required.
In general, a linear \ac{ECC} is a $K$-dimensional subspace of the full $N$-dimensional space of possible \ac{PUF} responses $\FF_2^N$. For this reason, the probability that $\xv^*$ is a valid codeword (or close to a valid codeword) is vanishingly small.
Therefore, the code space (or, equivalently, the \ac{PUF} response) must be \emph{offset} such that a codeword is close to the \ac{PUF} response.
This offset must be determined once and stored in non-volatile memory (fuses), in such a way that it does not leak any information about the secret.
There exist several implementations of such an offset.
We consider the commonly used \emph{fuzzy commitment} code-offset helper data algorithm \cite{delvaux2015helperdataalgos} as depicted in Fig.~\ref{fig:puf_overview}.

The operation of the \ac{PUF} is split into two phases: \emph{enrollment} and \emph{activation}.
At enrollment (e.g., in the factory), the golden \ac{PUF} response $\xv^*$ is determined, for example by querying the \ac{PUF} multiple times and performing majority voting. Moreover, a secret message $\mv \in \FF_2^K$ is either selected or randomly generated using a \ac{TRNG} and encoded into the codeword $\cv = \mv \Gm$.
Then, the helper data $\wv = \cv \oplus \xv^*$ is determined and stored in the fuses.
At activation (i.e., at every device readout), the noisy \ac{PUF} response $\xv$ is read out, and a vector close to the codeword $\yv = \xv \oplus \wv$ is obtained.
Since $\yv = \xv \oplus \wv = \xv \oplus \xv^* \oplus \cv = \cv \oplus \ev$, the codeword $\cv$ (or $\mv$) can be recovered by applying the \ac{ECC} decoder to $\yv$ to remove the error $\ev$, assuming the error is within the error-correction capability of the code.

\subsection{Polar Codes}
An $(N=2^n, K)$ Polar code $\Cc$ is constructed from the Polar transformation defined by the matrix $\Gm_N=\begin{bsmallmatrix}1&0\\1&1
\end{bsmallmatrix}^{\otimes n}$, where $(\cdot)^{\otimes n}$ denotes the $n$-fold application of the Kronecker product.
This transformation, along with a successive decoding schedule, results in a polarization effect, i.e., some channels become virtually noiseless and can carry uncoded information. The remaining channels are completely noisy. Hence they cannot carry information and are fixed to a known value (typically 0).
The Polar code is defined by selecting these $K$ reliable channels (or rows from the $\Gm_N$) called the information set $\Ic$, while the other $N-K$ row indices form the frozen set $\Fc$.
The synthetic channels exhibit a partial order \cite{bardet2016algebraicproperties} with respect to their reliability. Thus, a Polar code is fully specified by a few generators $\Ic_\mathrm{min}$, and the remaining elements of $\Ic$ are implied by the partial order \cite{geiselhart2021polaraut}.
Let $\mv\in \FF_2^K$ denote the payload (message) vector of $K$ information bits. The fully uncoded vector $\uv\in \FF_2^N$ is constructed from the information bits $\mv$ in the information positions $\Ic$ and zeros in the frozen positions $\Fc$. The codeword $\cv$ is obtained by computing $\cv = \uv \Gm_N$ using the arithmetic in $\FF_2$ (i.e., modulo 2).

\subsection{Successive Cancellation Decoding}
Polar codes can be decoded using the \ac{SC} algorithm \cite{schnabl1995rmgcc}. For each bit  $u_i$, \ac{SC} decoding determines the most likely value based on the received vector $\yv$ and the previously decoded bits $\hat{u}_0 \dots \hat{u}_{i-1}$,
where $\hat{u}_i = 0$ for frozen bits $i\in \mathcal{F}$.
\Ac{SC} decoding can be formulated recursively on the factor tree in \ac{LLR} domain. Let $\ellv$ denote the \ac{LLR} vector corresponding to the $\yv$.
First, $\ellv$ is split in two halves $\ellv = (\ellv_1 \mid \ellv_2) $ and the output \ac{LLR} for the first half is given as
\begin{equation*}\label{eq:scg}
	\ell_{1,i}' := f(\ell_{1,i}, \ell_{2,i}) := \sgn(\ell_{1,i})\sgn(\ell_{2,i}) \min \{ |\ell_{1,i}|, |\ell_{2,i}| \},
\end{equation*}
which approximates $\tilde{f}(a,b)= \ln {\frac{1+\operatorname{e}^{a+b}}{\operatorname{e}^{a}+\operatorname{e}^{b}}}$.
Then, $\ellv_1'$ is decoded further recursively, resulting in the estimate $\hat{\cv}_1'$. This is used to compute the output \ac{LLR} for the second half as
\begin{equation*}
	\ell_{2,i}' := g(\ell_{1,i}, \ell_{2,i}, \hat{c}_{1,i}') := (-1)^{\hat{c}_{1,i}'} \ell_{1,i} + \ell_{2,i},
\end{equation*}
and the decoding of $\ellv_2'$ is performed recursively, resulting in $\hat{\cv}_2'$. The overall codeword estimate is then computed as $\hat{\cv}=(\hat{\cv}_1'\oplus\hat{\cv}_2' \mid \hat{\cv}_2')$.
The recursion can be followed until the information bits $u_i$ are reached. However, usually, larger leaf nodes are reached early, which are simple to decode, such as Rate-0, Repetition (Rep), \ac{SPC}, and \mbox{Rate-1} codes  \cite{simplifiedSC}.
In the following, let $\operatorname{SC}(\yv)$ denote the decoding function that returns the codeword estimate $\hat{\cv}$ using the \ac{SC} decoding algorithm.

\subsection{Permutation Automorphism Group}
The (permutation) automorphism group of a code $\Cc$ is the set of permutations $\pi\in S_N$ that map each codeword onto another (not necessarily different) codeword, i.e.,
\begin{equation*}
	\operatorname{Aut}(\Cc) = \{\pi \mid \pi(\cv) \in \Cc \; \forall \cv \in \Cc\}.
\end{equation*}
The permutation automorphisms form a group under permutation composition, i.e., the following properties hold:
\begin{itemize}
	\item $\id \in \operatorname{Aut}(\Cc)$
	\item $\forall \pi, \sigma \in \operatorname{Aut}(\Cc)\!: \pi \circ \sigma \in \operatorname{Aut}(\Cc)$, where $(\pi \circ \sigma)(i) = \pi(\sigma(i))$
	\item $\forall \pi \in \operatorname{Aut}(\Cc)\!: \exists\pi^{-1} \in \operatorname{Aut}(\Cc)$ with $\pi \circ \pi^{-1} = \id$.
\end{itemize}
The $j$-fold application of a permutation $\pi$ (composition with itself) is denoted by $\pi^j$ and the \emph{order} $\ord(\pi)$ is the smallest positive number $j$ such that $\pi^j=\id$.

Polar codes exhibit affine permutation automorphisms of the form
\begin{equation*}
	\xv' = \Am \xv + \bv,
\end{equation*}
where $\xv$ and $\xv'$ are the \ac{LSB}-first binary representations\footnote{As an exception, we are using column vectors here.} of the codeword indices $i$ and $i'=\pi(i)$, respectively, $\bv \in \FF_2^n$ and $\Am \in \FF_2^{n\times n}$ is a block lower triangular invertible matrix with no non-zero elements above the block-diagonal given by a block profile $\sv$.
The block profile is uniquely defined by the information set $\Ic$ of the Polar code
\cite{geiselhart2021polaraut}. This group is called the \ac{BLTA} group denoted by $\operatorname{BLTA}(\sv)$.
For $\sv = [1,1,\dots,1]$, we have the \ac{LTA} group denoted by $\operatorname{LTA}(n)$.

\subsection{Automorphism Ensemble Decoding}
\Acf{AED} is a decoding algorithm that performs multiple decoding attempts on permuted versions of the noisy codeword $\yv$ \cite{geiselhart2021aedrm}.
To this end, a set of $M$ permutations $\pi_j \in \operatorname{Aut}(\Cc)$ is selected and the output of the $j$-th \ac{SC} decoding attempt is given by
\begin{equation*}
	\hat{\cv}_j = \pi_j^{-1}(\operatorname{SC}(\pi_j(\yv))).
\end{equation*}
Out of these codeword candidates, the most likely codeword is selected as the final codeword estimate
\begin{equation*}
	\hat{\cv} = \argmax_{0 \le j < M} p(\yv \mid \hat{\cv}_j).
\end{equation*}
In the following, \ac{SC}-based \ac{AED} with $M$ permutations is denoted by AE-SC-$M$.

In \cite{geiselhart2021aedrm}, it was shown that for all $\pi\in \operatorname{LTA}(n)$, $\pi_j^{-1}(\operatorname{SC}(\pi_j(\yv))) = \operatorname{SC}(\yv)$, i.e., the permutation is \emph{absorbed} by \ac{SC} decoding and there is no gain in using these permutations in \ac{AED}.
Therefore, it is sufficient to consider linear permutations whose $\Am$ matrix is a product of a permutation matrix $\Pm$ and an upper-triangular matrix $\Um$ bounded by the block profile $\sv$ \cite{geiselhart2021aedrm, bioglio2023groupproperties}.

For ease of description, we consider the slightly more general set of invertible block-diagonal matrices $\Am$ with block sizes $\sv=[s_0, s_1, \dots, s_{t-1}]$, called the \ac{BDL} group $\operatorname{BDL}(\sv)$.
Then $\Am \in \operatorname{BDL}(\sv)$ can be written as $\Am = \operatorname{diag}(\Am_0, \Am_1, \dots, \Am_{t-1})$, where $\Am_j \in \FF_2^{s_j \times s_j}$ and invertible.
As the sub-matrices $\Am_j$ do not interact and can be chosen independently, \ac{BDL} is the direct group product
\begin{equation*}
	\operatorname{BDL}(\sv) = \operatorname{GL}(s_0) \times \operatorname{GL}(s_1) \times \cdots \times \operatorname{GL}(s_{t-1})
\end{equation*}
where $\operatorname{GL}(m)$ denotes the general linear group of degree $m$ over $\FF_{2}$.
This leads to the following property about \ac{BDL} permutations.
\begin{theorem}
	Let $\Pm(\Am)$ denote the $N\times N$ permutation matrix of codeword bits corresponding to the linear transformation matrix $\Am$.
	For any block profile $\sv = [s_0, s_1, \dots, s_{t-1}]$, and $\Am = \operatorname{diag}(\Am_0, \Am_1, \dots, \Am_{t-1}) \in \operatorname{BDL}(\sv)$, we have
	\begin{equation*}
		\Pm(\Am) = \Pm(\Am_0) \otimes \Pm(\Am_1) \otimes \dots \otimes \Pm(\Am_{t-1}).
	\end{equation*}
\end{theorem}
\begin{IEEEproof}
	The statement follows from the fact that \ac{BDL} is a direct product of general linear groups \cite[Ch. 3.2]{serre1977representations}.
\end{IEEEproof}
\begin{example}
	Let $\sv=[2,2]$, $\Am_0 = \begin{bsmallmatrix}1&1\\0&1
	\end{bsmallmatrix}$ and $\Am_1 = \begin{bsmallmatrix}0&1\\1&0
\end{bsmallmatrix}$.
$\Am_0$ corresponds to the permutation $\pi_0 =[0,1,3,2]$ and $\Am_1$ to $\pi_1= [0,2,1,3]$.
The overall permutation corresponding to $\Am = \operatorname{diag}(\Am_0,\Am_1)$ is $\pi = [ 0,  1,  3,  2,  8,  9, 11, 10,  4,  5,  7,  6, 12, 13, 15, 14]$, where there are four blocks of four elements each, which are permuted corresponding to $\pi_0$ and $\pi_1$, respectively.

\end{example}

\section{Non-Uniform Quantization}
In contrast to wireless communications, where the decoder input $\yv$ is typically assumed to be the output of a Gaussian channel and of high resolution, the \ac{PUF} scenario is equivalent to a \ac{BSC} with bit-flip probability $\epsilon$, i.e., $\yv \in \FF_2^N$.
Assuming equally probable bits $c_i$, channel output \ac{LLR} of the $i$-th bit can be computed as
\begin{equation*}
	\ell_i = \ln \frac{P(y_i | c_i = 0)}{P(y_i | c_i = 1)} = (-1)^{y_i} \ln\frac{1-\epsilon}{\epsilon} .
\end{equation*}
Observe that \ac{SC} decoding is invariant to constant scaling of the input \acp{LLR}. Therefore, the common factor $\ln\tfrac{1-\epsilon}{\epsilon}$ can be dropped  %
and only a single bit of precision is required for each input \ac{LLR} $\ell_i \in \{\pm 1\}$ to the decoder.

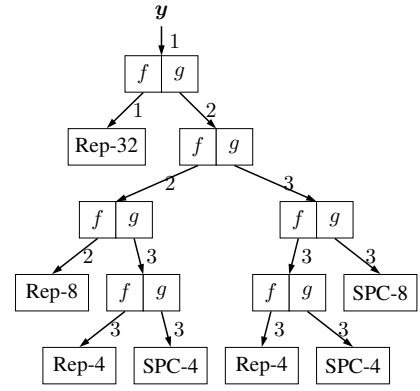
\begin{figure}
	\centering
	\resizebox{.615\linewidth}{!}{\input{fig/factor_tree}}
	\caption{Simplified \ac{SC} decoding factor tree of a $(64,17)$ Polar code with \ac{LLR} bit widths annotated; max. bit width is 3 bits.}
	\label{fig:factor_tree}
\end{figure}
As a consequence of the low input \ac{LLR} resolution, the decoding can be non-uniformly quantized (meaning different bit widths for different variables) without any loss in performance.
Let $\Vc(\ell)$ denote the set of different values that an \ac{LLR} $\ell$ may take. A binary digital implementation of \ac{SC} decoding requires $\lceil \log_2|\Vc(\ell)| \rceil$ bits to represent the value $\ell$. Observe that the $f$-function does not change the set $\Vc$, and hence does not increase the bit width to represent the outgoing \acp{LLR}. The $g$-function may output any of $\pm\ell_1 + \ell_2\;\forall \ell_1 \in \Vc(\ell_1), \ell_2 \in \Vc(\ell_2)$, at most doubling the size of $\Vc$. Hence, at most one bit more is required for the output than the inputs of the $g$-function. To reduce implementation complexity, we propose to limit the number of different values to a power of two, namely the maximum bit width of the decoder. For an integer-based realization of the \acp{LLR}, we divide the value by 2 if all elements of $\Vc$ are even. For example, after the first $g$-function with inputs $\pm 1$, the set of outputs is $\{-2,0,+2\}$, which may be scaled down to $\{-1,0,+1\}$, again due to the scaling invariance of \ac{SC} decoding.
Then, the values are clipped to $q$ bits, saturating at $\pm(2^{q-1}-1)$.

Fig.~\ref{fig:factor_tree} shows the simplified factor tree of a $(64,17)$ Polar code with binary channel output and a maximum quantization bit width of 3 bits. A large portion of the tree only requires 1 or 2 bits of quantization.
Furthermore, we observed that AE-\ac{SC} decoding is more robust to low-resolution \ac{LLR} quantizations when compared to pure SC decoding. Consequently, the proposed AE-\ac{SC} decoder can operate with 3-bit message quantizations even for deeper factor trees. Fig.~\ref{fig:bler_interleavers} shows an $(1024, 78)$ Polar code with 3-bit message quantization compared to the 5-bit \ac{SC} decoder. The corresponding 3-bit \ac{SC} decoder is not shown, but its performance significantly degrades.

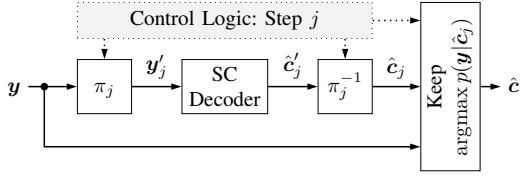
\begin{figure}
	\centering
	\resizebox{.8\linewidth}{!}{\input{fig/serial_aed}}
	\caption{Serial implementation of \ac{AED} with a single constituent decoder and switchable permutations.}
	\label{fig:serial_aed}
\end{figure}

\section{Serial Automorphism Ensemble Decoding}

\subsection{Serial Architecture}

\begin{figure}
	\begin{subfigure}{\columnwidth}
		\centering\scalebox{.718}{\input{fig/bigmux}}
		\caption{$M$ independent interleavers with ``big mux''}
		\label{fig:bigmux} \hfill
	\end{subfigure}
	\begin{subfigure}{\columnwidth}
		\centering\scalebox{.718}{\input{fig/cascaded_2}}
		\caption{$\log_2 M$ cascaded interleavers}
		\label{fig:cascaded} \hfill
	\end{subfigure}
	\begin{subfigure}{\columnwidth}
		\centering\scalebox{.718}{\input{fig/recursive}}
		\caption{A single recursive interleaver (simplified, registers omitted)}
		\label{fig:recursive}
	\end{subfigure}
	\caption{Different interleaver architectures to realize the permutations in a serial \ac{AED} implementation. The recursive de-interleaver requires multiple cycles which are not shown.}
	\label{fig:intl_arch}
\end{figure}
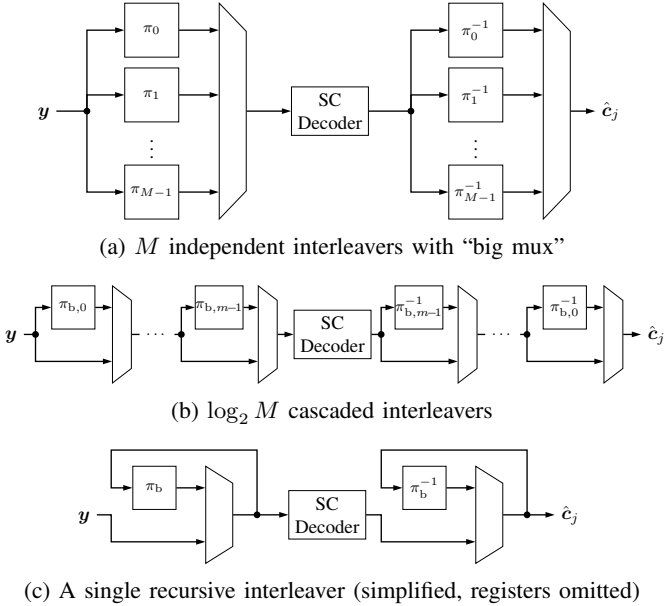

Compared to other ensemble decoding schemes, the component decoders in \ac{AED} are identical for each decoding attempt.
Therefore, for applications without stringent latency constraints, such as a \ac{PUF}, it becomes viable to re-use the same decoder block for each decoding attempt in a sequential fashion and only multiplex in a different permutation, as shown in Fig.~\ref{fig:serial_aed}.
Na\"ively, the permutations can be implemented by $M$ different, hard-wired interleavers, computing all $\yv_j$ in parallel, and then selecting the one required for the current decoding attempt using a big multiplexer (``big mux'').
This architecture is depicted in Fig.~\ref{fig:bigmux}.
Given a binary input signal, each of the two big multiplexers needs to switch $N$ parallel binary signals between $M$ inputs.
Note that, without loss of generality, one of the permutations may be the identity permutation.

\subsection{Cascaded Interleavers}

To reduce the complexity, we propose to select the permutations such that they are not completely independent anymore, but share common factors.

From $m$ base permutations $\pi_{\mathrm{b},i}$, $M=2^m$ permutations (including the identity permutation) are generated as
\begin{equation*}
	\pi_j = \prod_{i=0}^{m-1} \pi_{\mathrm{b},i}^{j_i},\label{eq:casc}
\end{equation*}
where $j_i$ denotes the $i$-th bit of the binary expansion of $j$.
A possible realization of these permutations in the serial \ac{AED} framework is depicted in Fig.~\ref{fig:cascaded}.
The resulting circuit involves $\log_2(M)$ two-input multiplexers (for $N$ parallel signals), which is in general much less complex than the single, $M$-input multiplexer. Moreover, only $\log_2(M)$ hard-wired interleavers are required, compared to the $M$ interleavers in the independent case.

\subsection{Recursive Interleavers}
To even further reduce the implementation complexity, it is also possible to generate multiple permutations (including the identity permutation) from a single base permutation as its powers
$
	\pi_j := \pi_{\mathrm{b}}^j %
$.
Formulated recursively, we have $\pi_j =   \pi_{\mathrm{b}} \circ \pi_{j-1}$, with $\pi_{0} = \id$.
Therefore, the permutations $\pi_j$ can be generated successively by routing the vector multiple times through an interleaver realizing $\pi_{\mathrm{b}}$, as shown in Fig.~\ref{fig:recursive}.
Note that while the permutation stage before decoding can output a new permutation every clock cycle, the de-permutation stage has to run $j$ times for the $j$-th permutation. %

This approach can generate up to $M=\operatorname{ord}(\pi_\mathrm{b})$ different permutations.
For $\pi_{\mathrm{b}}\in \operatorname{BDL}(\sv)$ with transformation matrix $\Am = \operatorname{diag}(\Am_0, \Am_1, \dots, \Am_{t-1})$, we have
\begin{equation*}
	\operatorname{ord}(\pi_\mathrm{b}) = \operatorname{LCM}\left(\operatorname{ord}(\Am_0),\operatorname{ord}(\Am_1),\dots, \operatorname{ord}(\Am_{t-1})\right),
\end{equation*}
which is maximized for $\Am_i$ being Singer cycles with $\operatorname{ord}(\Am_i) = 2^{s_i}-1$ \cite{hestenes1970singer}.

\subsection{Block Interleavers}
Independent of the realization of the permutations (e.g., cascaded or recursive), the size of the interleavers may be reduced by applying the same interleaver pattern multiple times to neighboring symbols (i.e., in a block-wise fashion).
While \ac{BDL} eliminates already many \ac{SC}-equivariant permutations from \ac{BLTA}, more permutations are absorbed by \ac{SC} decoding.
In particular, the block of size $s_0$ in the upper left corner of $\Am$ is absorbed if all size $2^{s_0}$ constituent codes are either rate-0, rate-1, repetition or \ac{SPC} codes, for which the \ac{SC} algorithm is \ac{ML}, and thus equivariant \cite{pillet2022classification}, \cite{ye2022completeinvariant}.
In these cases, one can set $\Am_0 = \Id_{s_0}$ without any loss in performance.

Even when they are non-equivariant, the \ac{AED} gain is smaller for small blocks in the top left of $\Am$ \cite{geiselhart2021polaraut}.
Therefore, to save implementation complexity, it may be beneficial to also set $\Am_0 = \Id_{s_0}$ in these cases.
The permutations move blocks of $b=2^{s_0}$ neighboring symbols based on a permutation $\sigma \in \operatorname{BDL}([s_1,\cdots, s_{t-1}])$, i.e.,
\begin{align*}
	\pi = \big[&\sigma(0),\sigma(0)+1,\dots,\sigma(0)+b-1, \\
				&\sigma(1),\sigma(1)+1,\dots,\sigma(1)+b-1, \quad \dots, \\
				&\sigma(N/b-1),\sigma(N/b-1)+1,\dots,\sigma(N/b-1)+b-1\big].
\end{align*}
\begin{example}
	Let $\sv=[2,2]$, with $\Am_1 = \begin{bsmallmatrix}0&1\\1&0
	\end{bsmallmatrix}$. We set $\Am_0 = \Id_2$.
	The overall permutation is $\pi = [ 0,  1,  2,  3,  8,  9, 10, 11,  4,  5,  6,  7, 12, 13, 14, 15]$, where blocks of size $b=4$ stay together.
\end{example}
Therefore, again at the cost of increased latency, the permutation $\pi$ can be realized by applying the small permutation $\sigma$ $b$ times in a bit-serial fashion.

\section{Results}\label{sec:results}
\subsection{Comparison of ECC Architectures}
We target a \ac{PUF} scenario with payload size of $K=312$ bits that should attain a maximum activation failure rate (\ac{BLER}) of $10^{-6}$ at worst-case physical bit-flip probability $\epsilon=0.22$.

\begin{figure}
	\centering
	\resizebox{\columnwidth}{!}{\input{fig/bler_comparison}}
	\caption{Comparison of \ac{BCH} and Polar-based coding schemes for $K \approx 312$.}
	\label{fig:bler_comparison}
\end{figure}
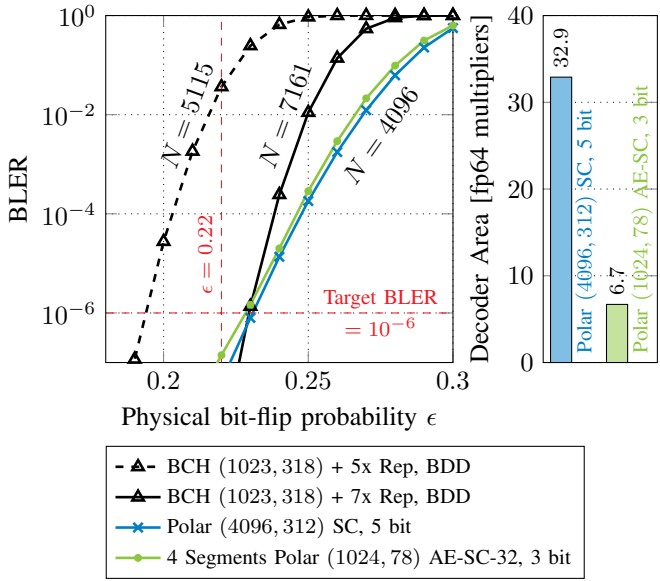

Fig.~\ref{fig:bler_comparison} compares the proposed Polar-\ac{AED} scheme with other \ac{ECC} architectures.
The best possible baseline system with \ac{BCH} and \ac{BDD} codes consists of an $(1023,318)$, $91$-error-correcting \ac{BCH} code and an inner 7-fold repetition code. Note that 5-fold repetition is not enough to meet the constraints.
Secondly, we consider a single $(4096,312)$ Polar code designed using \ac{DE} \cite{mori2009DE} for 5-bit \ac{SC} decoding.
By utilizing the more powerful \ac{AED} decoder with $M=32$ permutations, we can achieve the same performance using four segments of a $(1024, 78)$ Polar code and only 3-bit \ac{SC}, resulting in a decoder footprint that is five times smaller.
Areas are normalized to a pipelined double-precision floating-point multiplier, which serves as a proxy for a sizeable logic component.
Note that all segments must be correctly decoded, so the \ac{BLER} approximately quadruples compared to the single segment case.
The $(1024, 78)$ code was found using the symmetric partial order \cite{geiselhart2023ratecompatible} for the fixed block profile $\sv = [3, 7]$ and has minimum information set $\Ic_\mathrm{min} = \{255,505\}$.
All its $2^{s_0}=8$ bit \ac{SC} leaf node decoders are \ac{ML} decoders, and hence, bit permutations within the three least significant bits are equivariant.

The performance of all systems is roughly equal at the target operating point.
However, the \ac{BCH}+Repetition system requires codewords that are $1.75$ times as long (and, thus, $75\,\%$ more \ac{PUF} cells and fuses) as the Polar code based schemes.

\subsection{Permutation Selection and Realization}
Although a random selection of the permutations in \ac{AED} works well in practice, it can sometimes result in poorly performing ensembles.
Therefore, for each interleaver realization, we optimize the selection of the permutations using the greedy, data-driven approach proposed in \cite{kestel2023URLLC}.

\addtolength\abovecaptionskip{-10pt}
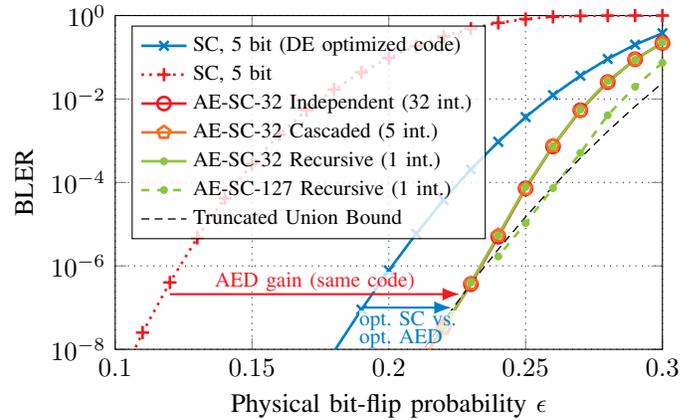
\begin{figure}
	\centering
	\input{fig/bler_1024}
	\caption{Error-rate performance results for different interleaver realizations for the $ (1024, 78) $ code with $\Ic_{\mathrm{min}}= [255,505]$.}
	\label{fig:bler_interleavers}
\end{figure}
\addtolength\abovecaptionskip{10pt}

Fig.~\ref{fig:bler_interleavers} compares different decoding approaches for the $(1024,78)$ code with $\Ic_{\mathrm{min}}=[255,505]$.
We consider \ac{SC} decoding with 5-bit quantization, which is approximately equal to full-precision decoding, and \ac{AED} with 3-bit quantization.
For the symmetric code design, \ac{AED} with $M=32$ achieves a \ac{BLER} of $10^{-6}$ at double the physical error rate, even with its coarser quantization.
Furthermore, the interleaver realizations show virtually identical performance, indicating no loss from the proposed cascaded or recursive interleaver designs.
Finally, as a baseline we show the best possible $(1024,78)$ Polar code for 5-bit \ac{SC} decoding obtained from \ac{DE}.
While this code performs better under \ac{SC} decoding, the symmetric code outperforms it with \ac{AED}.
The recursive interleaver uses a Singer cycle in the non-equivariant part, yielding up to $2^7-1=127$ decoding paths; it is the best-performing curve, and it closely follows the truncated union bound of the code.

\subsection{Interleaver Implementation Complexity}

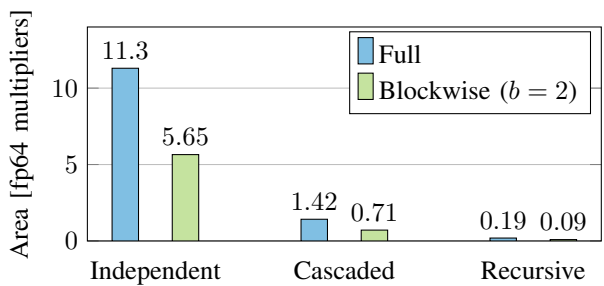
\begin{figure}
	\centering
	\input{fig/interleaver_area}
	\caption{Normalized hardware area of different interleaver realizations for $N=1024$ AE-SC-128. Area is normalized to a double-precision floating-point multiplier.}
	\label{fig:interleaver_area}
\end{figure}

We compare the hardware overheads of different interleavers using Verilog designs synthesized by the Synopsys toolchain with a modern 3\,nm industrial technology library.
Fig.~\ref{fig:interleaver_area} compares the post-synthesis area of a permutation unit of a $N=1024$ AE-SC-128 decoder using independent permutations (i.e., ``big mux'') to our optimized cascaded and recursive designs. Both a full and blockwise interleaver ($b=2$) are shown. The independent interleavers consume significant logic area, requiring the equivalent of $>10\times$ (non-blockwise) or $>5\times$ (blockwise) multipliers. The cascaded interleavers reduce the area to a more practical $<1.5$ multiplier-equivalents, while the recursive interleaver operates using only a single level of two-input multiplexing, reducing its area to a nominal size. The area of a multiplexer is roughly quadratic with its number of inputs, meaning that larger interleavers with more than 128 permutations would scale quadratically, while those with larger codeword sizes should scale linearly. At large codeword sizes or a large number of permutations, cascaded or recursive interleaving become imperative to avoid an explosion of complexity.

\section{Extension to Biased PUFs}

Ideally, \ac{PUF} cells are unbiased, meaning $x_i \sim \operatorname{Ber}(p)$ with $p=0.5$ across devices.
However, if $p\ne 0.5$,  the asymmetric tendency towards 0 or 1 will indicate a global bias and
the entropy of $\xv$ will be reduced and helper data $\wv$ may leak information about $\xv^*$ \cite{maes2016biasedpufs}.
In such cases, debiasing is a remedy  to maintain system security.
As depicted in Fig.~\ref{fig:puf_overview}, debiasing is applied at the output of the \ac{PUF} cells.
An efficient debiasing scheme is \ac{VNPO} \cite{maes2016biasedpufs}, which considers \ac{PUF} cell pairs. At enrollment, pairs with identical values are discarded and kept pair positions are stored in debiasing helper data $\dv$.
If one of the bits of the kept pairs is flipped, an implicit two-fold repetition code will be formed.
Moreover, larger inner repetition codes may be beneficial to obtain low-rate codes with small decoder footprints.
When the corresponding values are combined, higher-precision \acp{LLR} will be formed that can be naturally processed by Polar decoders.
Our proposed \ac{AED} scheme can therefore be straightforwardly applied to \acp{PUF} with debiasing when the input bit width is increased accordingly.
Note that the concept of block interleavers can also be applied here, either for the bits of the independent observations which are combined after interleaving, or the quantized \ac{LLR} representations if combined before interleaving.

\section{Conclusion}\label{sec:conclusion}

In this work, we proposed a novel serial \ac{AED} scheme for Polar codes combining competitive error rate performance with high area efficiency. Our approach introduces cascaded and recursive interleavers that efficiently scale the number of \ac{AED} candidates without requiring expensive large multiplexers, enabling practical hardware implementation. The proposed aggressive 3-bit quantization strategy significantly reduces decoder complexity while maintaining near-\ac{ML} decoding performance, demonstrating the robustness of \ac{AED} to quantization effects. Our coding scheme achieves the same $10^{-6}$ block error rate as \ac{BCH}-based baselines while requiring 1.75$\times$ fewer codeword bits, directly translating to reduced helper data storage and chip area. Contrary to classical \ac{BCH} decoders, the proposed scheme naturally supports soft-decoding, opening a plethora of potential use cases.

\bibliographystyle{IEEEtran}
\bibliography{references}

\end{document}

%% file: acronyms.tex
\begin{acronym}
\acro{3GPP}{3rd Generation Partnership Project}
\acro{5G}{fifth generation mobile telecommunication}
\acro{AED}{automorphism ensemble decoding}
\acro{AE-SC}{automorphism ensemble successive cancellation}
\acro{AWGN}{additive white Gaussian noise}
\acro{ASIC}{application-specific integrated circuit}
\acro{BDD}{bounded distance decoding}
\acro{BDL}{block diagonal linear}
\acro{BCH}{Bose--Ray-Chaudhuri--Hocquenghem}
\acro{BEC}{binary erasure channel}
\acro{BER}{bit error rate}
\acro{BI-DMC}{binary input discrete memoryless channel}
\acro{BLER}{block error rate}
\acro{BLTA}{block lower-triangular affine}
\acro{BMI}{bitwise mutual information}
\acro{BP}{belief propagation}
\acro{BPL}{belief propagation list}
\acro{BPSK}{binary phase shift keying}
\acro{BSC}{binary symmetric channel}
\acro{CA-SCL}{CRC-aided successive cancellation list}
\acro{CN}{check node}
\acro{CP}{Chase-Pyndiah}
\acro{CRC}{cyclic redundancy check}
\acro{CSI}{Channel State Information}
\acro{DBT}{dichotomous binary tree}
\acro{DE}{density evolution}
\acro{DMC}{discrete memoryless channel}
\acro{DR}{delta recovery}
\acro{eMBB}{enhanced mobile broadband}
\acro{ECC}{error-correcting code}
\acro{FER}{frame error rate}
\acro{FFG}{Forney-style factor graph}
\acro{FHT}{fast Hadamard transform}
\acro{FSSCL}[Fast-SSCL]{fast simplified successive cancellation list}
\acro{GenAlg}{genetic algorithm}
\acro{GRAND}{guessing random additive noise decoding}
\acro{HD}{hard decision}
\acro{HDD}{hard decision decoding}
\acro{iBDD}{iterative bounded-distance decoding}
\acro{IBE}{information bit extraction}
\acro{LDPC}{low-density parity-check}
\acro{LLR}{log-likelihood ratio}
\acro{LSB}{least significant bit}
\acro{LTA}{lower-triangular affine}
\acro{MAP}{maximum a posteriori}
\acro{MBBP}{multiple-bases belief propagation}
\acro{MI}{mutual information}
\acro{ML}{maximum likelihood}
\acro{MSB}{most significant bit}
\acro{MWPC-BP}{minimum-weight parity-check BP}
\acro{OSD}{ordered statistic decoding}
\acro{ORBGRAND}{ordered reliability bits GRAND}
\acro{ORDEPT}{ordered reliability direct error pattern testing}
\acro{PAC}{polarization-adjusted convolutional}
\acro{PAR}{placement and routing}
\acro{PC}{product code}
\acro{PDBT}{perfect dichotomous binary tree}
\acro{PFT}{polar factor tree}
\acro{PL}{permutation linear}
\acro{PM}{path metric}
\acro{PPV}{Polyanskyi--Poor--Verd'{u}}
\acro{PTPC}{pre-transformed polar code}
\acro{PUF}{physical unclonable function}
\acro{PVT}{process, voltage and temperature}
\acro{QC}{quasi-cyclic}
\acro{REP}{repetition}
\acro{RM}{Reed--Muller}
\acro{RS}{Reed--Solomon}
\acro{RREF}{reduced row echelon form}
\acro{SCAN}{soft cancellation}
\acro{SC}{successive cancellation}
\acro{SCAL}{successive cancellation automorphism list}
\acro{SCL}{successive cancellation list}
\acro{SD}{soft decision}
\acro{SDD}{soft decision decoding}
\acro{SGD}{stochastic gradient descent}
\acro{SISO}{soft-in/soft-out}
\acro{SRAM}{static random access memory}
\acro{TRNG}{true random number generator}
\acro{SNR}{signal-to-noise-ratio}
\acro{SPC}{single parity check}
\acro{URLLC}{ultra-reliable low-latency communications}
\acro{UTL}{upper-triangular linear}
\acro{VN}{von Neumann}
\acro{VNPO}{von Neumann with pair output}
\acro{VLSI}{very-large-scale integration}
\end{acronym}

%% file: colours.tex
\definecolor{mittelblau}{RGB}{0, 126, 198}
\definecolor{violettblau}{cmyk}{0.9, 0.6, 0, 0}
\definecolor{rot}{RGB}{238, 28 35}
\definecolor{apfelgruen}{RGB}{140, 198, 62}
\definecolor{gelb}{RGB}{255, 229, 0}
\definecolor{orange}{RGB}{244, 111, 33}
\definecolor{pink}{RGB}{237, 0, 140}
\definecolor{lila}{RGB}{128, 10, 145}
\definecolor{hellgrau}{RGB}{224, 224, 224}
\definecolor{mittelgrau}{RGB}{128, 128, 128}
\definecolor{dunkelgrau}{RGB}{80,80,80}
\definecolor{anthrazit}{RGB}{19, 31, 31}
\definecolor{aqua}{RGB}{0, 255, 255}

%% file: macros.tex
\renewcommand{\vec}[1]{\boldsymbol{#1}}
\newcommand{\vecs}[1]{\boldsymbol{#1}}

\newcommand{\bv}{\vec{b}}
\newcommand{\cv}{\vec{c}}
\newcommand{\dv}{\vec{d}}
\newcommand{\ev}{\vec{e}}

\newcommand{\mv}{\vec{m}}

\newcommand{\sv}{\vec{s}}

\newcommand{\uv}{\vec{u}}

\newcommand{\wv}{\vec{w}}
\newcommand{\xv}{\vec{x}}
\newcommand{\yv}{\vec{y}}

\newcommand{\ellv}{\vecs{\ell}}

\newcommand{\Am}{\vec{A}}

\newcommand{\Gm}{\vec{G}}

\newcommand{\Id}{\vec{I}}

\newcommand{\Pm}{\vec{P}}

\newcommand{\Um}{\vec{U}}

\newcommand{\Cc}{{\cal C}}

\newcommand{\Fc}{{\cal F}}

\newcommand{\Ic}{{\cal I}}

\newcommand{\Sc}{{\cal S}}

\newcommand{\Vc}{{\cal V}}

\newcommand{\FF}{\mathbb{F}}

\renewcommand{\hbar}[1]{\mathbb{#1}}   %

\newcommand{\id}{\mbox{id}}

\newcommand{\sgn}{\operatorname{sgn}}

\DeclareMathOperator*\ord{ord}

\newcommand{\argmax}{\mathop{\mathrm{argmax}}}

\newcolumntype{?}{!{\vrule width 1pt}}

%% file: fig/fuzzy_commitment.tex
\begin{tikzpicture}
	
	\tikzstyle{dspline} = [line width = 0.25mm]
	
	\tikzstyle{arrow} = [line width = 0.25mm, -Stealth]
	
	\tikzstyle{block} = [
	draw, 
	fill=white,
	align=center,
	minimum height=0.9cm,
	minimum width=2cm,
	]
	\tikzstyle{cell} = [
	draw, 
	fill=white,
	align=center,
	minimum height=0.9cm,
	minimum width=0.9cm,
	]
	
	\tikzstyle{mem} = [
	dashed, 
	rounded corners=5pt,
	fill=white!90!gray,
	align=center,
	minimum width=1.8cm,
	]
	
	\draw[loosely dashed, gray, line width=1pt] (-.5,0) -- (12.7,0);
	
	\node[rotate=90,gray] at (-.4,1.5) {Enrollment};
	\node[rotate=90,gray] at (-.4,-1.5) {Activation};

	\node[cell] (puf1) at (0.5,1.5) {PUF\\Cells};
	\node[block, fill=yellow, dashed] (deb1) at (3,1.5) {Debiasing};
	\node[dspadder] (add1) at (5,1.5) {};
	\node[block] (enc) at (7,1.5) {ECC\\Encoder};
	\node[align=center] (trng) at (8.5,2.3) {TRNG Secret\\[-3pt]$\mv$};
	\node[dspnodefull] (split) at (8.5,1.5) {};
	\node[block,dashed] (hacc1) at (10,1.5) {Entropy\\Accumulator};
	\node[align=center] (out1) at (12.2,1.5) {PUF\\Output};
	
	\node[mem, 	fill=yellow!40!white,] (debdata) at (3,0) {\footnotesize Debiasing\\\footnotesize Helper Data};
	\node[mem] (helperdata) at (5,0) {\footnotesize ECC\\\footnotesize Helper Data};
	
	\draw[dspconn] (puf1) -- node[midway, above]{$\tilde{\xv}\mathstrut$} (deb1);
	\draw[dspconn] (deb1) -- node[midway, above]{$\xv^*\mathstrut$} (add1);
	\draw[dspline] (trng) -- (split); %
	\draw[dspconn] (split) -- (enc);
	\draw[dspconn] (enc) -- node[midway, above]{$\cv\mathstrut$} (add1);
	\draw[dspconn,densely dotted] (split) -- (hacc1);
	\draw[dspconn,densely dotted] (hacc1) -- (out1);
	\draw[dspconn, densely dotted] (deb1) -- node[midway, right]{$\dv$} (debdata);
	\draw[dspconn] (add1) -- node[midway, right]{$\wv$} (helperdata);

	\node[cell] (puf2) at (0.5,-1.5) {PUF\\Cells};
	\node[block, fill=yellow, dashed] (deb2) at (3,-1.5) {Debiasing};
	\node[dspadder] (add2) at (5,-1.5) {};
	\node[block] (dec) at (7,-1.5) {ECC\\Decoder};
	\node[block] (hacc2) at (10,-1.5) {Entropy\\Accumulator};
	\node[align=center] (out2) at (12.2,-1.5) {PUF\\Output};
	
	\draw[dspconn, densely dotted] (debdata) -- node[midway, right]{$\dv$} (deb2);
	\draw[dspconn] (helperdata) -- node[midway, right]{$\wv$} (add2);
	
	\draw[dspconn] (puf2) -- node[midway, above]{$\tilde{\xv}\mathstrut$} (deb2);
	\draw[dspconn] (deb2) -- node[midway, above]{$\xv\mathstrut$} (add2);
	\draw[dspconn] (add2) -- node[midway, above]{$\yv\mathstrut$} (dec);
	\draw[dspconn] (dec) -- node[midway, above]{$\hat{\mv}\mathstrut$} (hacc2);
	\draw[dspconn] (hacc2) -- (out2);
	
\end{tikzpicture}

%% file: fig/factor_tree.tex
\begin{tikzpicture}[scale=.6]
	
	\tikzstyle{normalnode} = [dspnodefull,minimum size=1mm]
	\tikzstyle{dspline} = [line width = 0.25mm]
	
	\tikzstyle{arrow} = [dspconn]
	
	\tikzstyle{split} = [
	draw, 
	fill=white,
	minimum height=.6cm,
	minimum width=.6cm,
	]
	
	\tikzstyle{leaf} = [
	draw, 
	fill=white,
	minimum height=.6cm,
	minimum width=1.2cm,
	]
	\tikzset{
		fgpair/.pic={			
			\node[split] (f) at (-.3,0) {$f$};
			\node[split] (g) at (+.3,0) {$g$};
			
			\coordinate (-in)  at (0,0.3); %
			\coordinate (-f)   at (f.south);
			\coordinate (-g)   at (g.south);
		},
	}
	\node (y) at (-1.5,1.7) {$\yv$};
	\pic (fg0) at (-1.5,0) {fgpair};
	\node[leaf] (r1) at (-3,-2) {Rep-32};
	\pic (fg1) at (0,-2) {fgpair};
	\pic (fg2) at (-2.75,-4) {fgpair};
	\pic (fg3) at (2.75,-4) {fgpair};
	\node[leaf] (r2) at (-4.5,-6) {Rep-8};
	\pic (fg4) at (-2,-6) {fgpair};
	\pic (fg5) at (2,-6) {fgpair};
	\node[leaf] (s1) at (4.5,-6) {SPC-8};
	\node[leaf] (r3) at (-3.75,-8) {Rep-4};
	\node[leaf] (s2) at (-1.25,-8) {SPC-4};
	\node[leaf] (r4) at (1.25,-8) {Rep-4};
	\node[leaf] (s3) at (3.75,-8) {SPC-4};
	
	\draw[arrow] (y) -- node[midway,right]{$1$} (fg0-in);
	\draw[arrow] (fg0-f) -- node[midway,right]{$1$} (r1.north);
	\draw[arrow] (fg0-g) -- node[midway,right]{$2$} (fg1-in);
	\draw[arrow] (fg1-f) -- node[midway,right]{$2$} (fg2-in);
	\draw[arrow] (fg1-g) -- node[midway,right]{$3$} (fg3-in);
	\draw[arrow] (fg2-f) -- node[midway,right]{$2$} (r2.north);
	\draw[arrow] (fg2-g) -- node[midway,right]{$3$} (fg4-in);
	\draw[arrow] (fg3-f) -- node[midway,right]{$3$} (fg5-in);
	\draw[arrow] (fg3-g) -- node[midway,right]{$3$} (s1.north);
	\draw[arrow] (fg4-f) -- node[midway,right]{$3$} (r3.north);
	\draw[arrow] (fg4-g) -- node[midway,right]{$3$} (s2.north);
	\draw[arrow] (fg5-f) -- node[midway,right]{$3$} (r4.north);
	\draw[arrow] (fg5-g) -- node[midway,right]{$3$} (s3.north);
	
\end{tikzpicture}

%% file: fig/serial_aed.tex
\begin{tikzpicture}
	\tikzset{
		edge/.style = {thick,black},
		decrect/.style={rectangle, draw, minimum height=.9cm, fill=white},
		intrect/.style={rectangle, draw, minimum size=.9cm, fill=white}
	}
	
	\tikzstyle{conn} = [dspconn];
	
	\node[] (Lch) at (-.5,0) {$\yv\mathstrut$};
	\node[dspnodefull] (split) at (0,0){};	
	\node[intrect] (int) at (1, 0) {$ \pi_j \mathstrut$};
	
	\node[decrect, align=center] (d) at (3,0) {SC\\Decoder};
	
	\coordinate (h1) at (3.4,0.7) {};
	
	\node[intrect] (deint) at (5, 0) {$ \pi_j^{-1}\mathstrut $};
	
	\node[rectangle, draw, minimum width=2.8cm, minimum height=0.3cm, align=center,rotate=90, fill=white] (decide) at (6.75,0)
	{Keep\\$\operatorname{argmax} p(\yv|\hat{\cv}_j)$};
	
	\node[] (end) at (7.8,0) {$\hat{\cv}\mathstrut$};
	\coordinate (help) at (0,-1) {};
	
	\node[rectangle, draw, dotted, minimum width=4.9cm, minimum height=0.6cm,align=center, fill=white!90!gray] (ctrl) at (3,1.1) {Control Logic: Step $j$};
	
	\draw [edge,conn, dotted] (int.north|-ctrl.south) -- (int);
	\draw [edge,conn, dotted] (deint.north|-ctrl.south) -- (deint);
	\draw [edge,conn, dotted] (ctrl) -- (decide.north|-ctrl.east);
	
	\draw [edge,conn] (Lch)--(split)-- (int);
	
	\draw [edge,conn](int) to node[above] {$\yv'_j\mathstrut$} (d);
	
	\draw [edge,conn](d) to node[above] {$\hat{\cv}'_j\mathstrut$} (deint);
	
	\draw [edge,conn](deint) to node[above] {$\hat{\cv}_j\mathstrut$} (decide.north|-deint.east);
	
	\draw [edge,conn] (split)|-(help)--(decide.north|-help.east);
	
	\draw [edge,conn](decide.south|-end.west) -- (end);

\end{tikzpicture}

%% file: fig/bigmux.tex
\begin{tikzpicture}
	
	\tikzstyle{normalnode} = [dspnodefull,minimum size=1mm]
	\tikzstyle{dspline} = [line width = 0.25mm]
	
	\tikzstyle{arrow} = [line width = 0.25mm, -Stealth]
	
	\tikzstyle{dec} = [
	draw, 
	fill=white,
	]
	
	\tikzstyle{perm} = [
	draw, 
	fill=white,
	minimum size=1cm,
	inner sep=0pt,
	]

	\tikzstyle{mux} = [
	trapezium,
	trapezium stretches = true,
	minimum height = .5cm,
	rotate=-90,
	draw, 
	fill=white,
	]
	
	\node (y) at (0,0) {$\yv\mathstrut$};
	\node[dspnodefull] (split) at (0.8,0) {};
	\draw[dspline] (y) -- (split);
	
	\node[perm] (pi0) at (2, 1.5) {\footnotesize $\pi_0$};
	\node[perm] (pi1) at (2, 0.3) {\footnotesize$\pi_1$};
	\node[] 		  at (2,-0.6) {$\vdots$};
	\node[perm] (pi2) at (2,-1.5) {\footnotesize $\pi_{M-1}$};
	
	\node[mux, minimum width=4cm, trapezium angle = 80] (mux1) at (3.5,0) {};
	
	\draw[dspconn] (split) |- (pi0);
	\draw[dspconn] (split) |- (pi1);
	\draw[dspconn] (split) |- (pi2);
	\draw[dspconn] (pi0) -- (pi0 -| mux1.south);
	\draw[dspconn] (pi1) -- (pi1 -| mux1.south);
	\draw[dspconn] (pi2) -- (pi2 -| mux1.south);
			
	\node[dec, align=center] (dec) at (5.3,0) {SC\\Decoder};
	\draw[dspconn] (mux1) -- (dec);
	
	\node[dspnodefull] (split2) at (6.8,0) {};
	\draw[dspline] (dec) -- (split2);
	
	\node[perm] (pi0i) at (8, 1.5) {\footnotesize $\pi_0^{-1}$};
	\node[perm] (pi1i) at (8, 0.3) {\footnotesize$\pi_1^{-1}$};
	\node[] 		   at (8,-0.6) {$\vdots$};
	\node[perm] (pi2i) at (8,-1.5) {\footnotesize $\pi_{M-1}^{-1}$};
	
	\node[mux, minimum width=4cm, trapezium angle = 80] (mux2) at (9.5,0) {};
	
	\draw[dspconn] (split2) |- (pi0i);
	\draw[dspconn] (split2) |- (pi1i);
	\draw[dspconn] (split2) |- (pi2i);
	\draw[dspconn] (pi0i) -- (pi0i -| mux2.south);
	\draw[dspconn] (pi1i) -- (pi1i -| mux2.south);
	\draw[dspconn] (pi2i) -- (pi2i -| mux2.south);

	\node (chat) at (10.5,0) {$\hat{\cv}_j\mathstrut$};
	\draw[dspconn] (mux2) -- (chat);

\end{tikzpicture}

%% file: fig/cascaded_2.tex
\begin{tikzpicture}
	
	\tikzstyle{normalnode} = [dspnodefull,minimum size=1mm]
	\tikzstyle{dspline} = [line width = 0.25mm]
	
	\tikzstyle{arrow} = [line width = 0.25mm, -Stealth]
	
	\tikzstyle{dec} = [
	draw, 
	fill=white,
	]
	
	\tikzstyle{perm} = [
	draw, 
	fill=white,
	minimum size=.8cm,
	inner sep=0pt,
	]
	
	\tikzstyle{mux} = [
	trapezium,
	trapezium stretches = true,
	minimum height = .35cm,
	rotate=-90,
	draw, 
	fill=white,
	]
	
	\tikzset{
		pics/stage/.style args={#1}{
			code={
				\node[dspnodefull] (split) at (0,0) {};
				\node[perm] (p) at (.7, .5) {\footnotesize #1};
				\coordinate (help) at (.7,-.5) {};
				\node[mux, minimum width=1.8cm, trapezium angle = 78] (mux) at (1.6,0) {};
				
				\draw[dspconn] (split) |- (p);
				\draw[dspconn] (p) -- (p -| mux.south);
				\draw[dspconn] (split) |- (help -| mux.south);
				
				\coordinate (-in) at (split);
				\coordinate (-out) at (mux.north);
				\coordinate (-sel) at (mux.west);
		}}
	}
	
	\node (input) at (-0.45,0) {$\yv\mathstrut$};
	
	\pic (s1) at (0,0) {stage={$\pi_{\mathrm{b},0}$}};
	\pic (s3) at (2.7,0) {stage={$\pi_{\mathrm{b},m\!-\!1}$}};
	\node[] (dots1) at (2.25,0) {\footnotesize$\cdots$};
	\draw[dspline] (input) -- (s1-in);
	\draw[dspline] (s1-out) -- (dots1);
	\draw[dspline] (dots1) -- (s3-in);
	
	\node[dec, align=center] (dec) at (5.5,0) {SC\\Decoder};
	\draw[dspconn] (s3-out) -- (dec);
	
	\pic (s4) at (6.4,0) {stage={$\pi_{\mathrm{b},m\!-\!1}^{-1}$}};
	\pic (s6) at (9.1,0) {stage={$\pi_{\mathrm{b},0}^{-1}$}};
	\node[] (dots2) at (8.65,0) {\footnotesize$\cdots$};
	\draw[dspline] (dec) -- (s4-in);
	\draw[dspline] (s4-out) -- (dots2);
	\draw[dspline] (dots2) -- (s6-in);

	\node (chat) at (11.5,0) {$\hat{\cv}_j\mathstrut$};
	\draw[dspconn] (s6-out) -- (chat);

\end{tikzpicture}

%% file: fig/recursive.tex
\begin{tikzpicture}
	
	\tikzstyle{normalnode} = [dspnodefull,minimum size=1mm]
	\tikzstyle{dspline} = [line width = 0.25mm]
	
	\tikzstyle{arrow} = [line width = 0.25mm, -Stealth]
	
	\tikzstyle{dec} = [
	draw, 
	fill=white,
	]
	
	\tikzstyle{perm} = [
	draw, 
	fill=white,
	minimum size=.8cm,
	inner sep=0pt,
	]
	
	\tikzstyle{mux} = [
	trapezium,
	trapezium stretches = true,
	minimum height = .5cm,
	rotate=-90,
	draw, 
	fill=white,
	]
	
	\node (y) at (0,0) {$\yv\mathstrut$};
	\node[perm] (pi0) at (1.3, 0.5) {\footnotesize $\pi_\mathrm{b}$};
	\coordinate (help) at (0.5,-.5) {};
	\node[mux, minimum width=1.8cm, trapezium angle = 75] (mux1) at (2.5,0) {};
	\draw[dspconn] (pi0) -- (pi0 -| mux1.south);
	\draw[dspconn] (y) -| (help) -- (help -| mux1.south);
	
	\node[dspnodefull] (feedback) at (3.2,0) {};
	\node[dec, align=center] (dec) at (4.5,0) {SC\\Decoder};
	\draw[dspconn] (mux1.north) -- (feedback) -- (dec);
	\coordinate (help2) at (0.5,1.2) {};
	\draw[dspconn] (feedback) |- (help2) |- (pi0);
	
	\node[perm] (pi1) at (6.3, 0.5) {\footnotesize $\pi_\mathrm{b}^{-1}$};
	\coordinate (help3) at (5.5,-.5) {};
	\node[mux, minimum width=1.8cm, trapezium angle = 75] (mux2) at (7.5,0) {};
	\draw[dspconn] (pi1) -- (pi1 -| mux2.south);
	\draw[dspconn] (dec) -| (help3) -- (help3 -| mux2.south);
	
	\node (chat) at (9,0) {$\hat{\cv}_j\mathstrut$};	
	\node[dspnodefull] (feedback2) at (8.2,0) {};
	\draw[dspconn] (mux2.north) -- (feedback2) -- (chat);
	\coordinate (help4) at (5.5,1.2) {};
	\draw[dspconn] (feedback2) |- (help4) |- (pi1);

\end{tikzpicture}

%% file: fig/bler_comparison.tex
\begin{tikzpicture}
		\begin{groupplot}[
			group style={
				group size=2 by 1, %
				horizontal sep=1.2cm %
			},
			width=\linewidth,
			height=.7\linewidth,
			grid style={dotted,anthrazit},			
			yminorticks=true,
			ymajorgrids
			]
		
		\nextgroupplot[
		width=.7\linewidth,
		xmajorgrids,
		legend columns=1,
		legend style={at={(0,-.24)},anchor=north west},
		legend cell align={left},
		xlabel={Physical bit-flip probability $\epsilon$},
		ylabel={BLER},
		y label style={at={(axis description cs:-0.2,.5)},anchor=south},
		legend image post style={mark indices={}},
		ymode=log,
		mark size= 2.5pt,
		xmin=0.18,
		xmax=0.3,
		ymin=1e-7,
		ymax=1,
		]

		\addplot[color=black, dashed, mark=triangle, mark options={solid}, line width = 1pt]
		table[col sep=comma]{
		0.300,1.000e+00
		0.290,1.000e+00
		0.280,1.000e+00
		0.270,9.999e-01
		0.260,9.950e-01
		0.250,9.326e-01
		0.240,6.579e-01
		0.230,2.454e-01
		0.220,3.648e-02
		0.210,1.824e-03
		0.200,2.759e-05
		0.190,1.165e-07
		0.180,1.272e-10
				};
	\label{plot:long_polar}
\addlegendentry{\footnotesize BCH $(1023,318)$ + 5x Rep, BDD};
		
		\addplot[color=black, solid, mark=triangle, mark options={solid}, line width = 1pt]
		table[col sep=comma]{
		0.300,9.999e-01
		0.290,9.936e-01
		0.280,9.026e-01
		0.270,5.411e-01
		0.260,1.376e-01
		0.250,1.107e-02
		0.240,2.437e-04
		0.230,1.357e-06
		0.220,1.813e-09
		0.210,5.552e-13
		};
	\label{plot:long_polar}
	\addlegendentry{\footnotesize BCH $(1023,318)$ + 7x Rep, BDD};
		
		\addplot[color=mittelblau, solid, mark=x, mark options={solid}, line width = 1pt]
		table[col sep=comma]{
		0.200,1.256e-10
		0.210,2.208e-09
		0.220,4.174e-08
		0.230,7.967e-07
		0.240,1.360e-05
		0.250,1.815e-04
		0.260,1.763e-03
		0.270,1.231e-02
		0.280,6.235e-02
		0.290,2.266e-01
		0.300,5.591e-01
		};
		\label{plot:long_polar}
		\addlegendentry{\footnotesize Polar $(4096,312)$ SC, 5 bit};
				
		\addplot[color=apfelgruen, solid, mark=*, mark options={solid}, line width = 1pt, mark size=1pt]
		table[col sep=comma]{
		0.300,6.316e-01
		0.290,3.129e-01
		0.280,9.787e-02
		0.270,2.135e-02
		0.260,2.926e-03
		0.250,2.880e-04
		0.240,2.008e-05
		0.230,1.464e-06
		0.220,1.408e-07
		0.210,1.256e-08
		};
		\label{plot:indep}
		\addlegendentry{\footnotesize 4 Segments Polar $(1024,78)$ AE-SC-32, 3 bit};
		
	\node[rotate=70] at (axis cs: .208, 1e-02) {$N=5115$} ;
	\node[rotate=70] at (axis cs: .242, 1e-02) {$N=7161$} ;
	\node[rotate=55] at (axis cs: .275, 4e-03) {$N=4096$} ;
	
	\draw[color=rot, dashed] 
	(axis cs:0.18, 1e-6) -- (axis cs:0.3, 1e-6);
	\draw[color=rot, dashed] 
	(axis cs:0.22, 1e-7) -- (axis cs:0.22, 1);
	
	\node[rotate=90, anchor=south, rot] at (axis cs: .22, 1.5e-5) {\footnotesize $\epsilon=0.22$} ;
	\node[anchor=center, rot, align=center] at (axis cs: .275, 1e-6) {\footnotesize Target BLER \\\footnotesize $= 10^{-6}$ } ;
		
	\nextgroupplot[
		width=.35\linewidth,
		ybar,
		enlarge x limits=0.35,
		ylabel={Decoder Area [fp64 multipliers]},
		y label style={at={(axis description cs:-0.35,.5)},anchor=south},
		ymin=0, ymax=40,
		xmax=2.3,
		xmin=1.1,
    	xtick=\empty,        %
		xticklabels=\empty,  %
		nodes near coords, 
		every node near coord/.append style={
			rotate=90, anchor=west, font=\footnotesize
		},
		every axis plot/.append style={
			ybar,
			bar width=8pt,
			bar shift=0pt,
			fill
		}
	]
	\addplot[black, fill=mittelblau!50] coordinates {(1,32.9)};
	\addplot[black, fill=apfelgruen!50] coordinates {(2,6.7)};
	\node[rotate=90, anchor=west, mittelblau] at (axis cs: 1.45,0) {\footnotesize Polar $(4096,312)$ SC, 5 bit};
	\node[rotate=90, anchor=west, apfelgruen] at (axis cs: 2.45,0) {\footnotesize Polar $(1024,78)$ AE-SC, 3 bit};
	 
\end{groupplot}
\end{tikzpicture}

%% file: fig/bler_1024.tex
\begin{tikzpicture}
	\begin{axis}[
		width=\linewidth,
		height=.68\linewidth,
		grid style={dotted,anthrazit},
		xmajorgrids,
		yminorticks=true,
		ymajorgrids,
		legend columns=1,
		legend pos=north west,   
		legend cell align={left},
		legend style={fill,fill opacity=0.8,text opacity=1},
		xlabel={Physical bit-flip probability $\epsilon$},
		ylabel={BLER},
		legend image post style={mark indices={}},
		ymode=log,
		ytick= {1,1e-2,1e-4,1e-6,1e-8},
		mark size= 2.5pt,
		xmin=0.1,
		xmax=0.3,
		ymin=1e-08,
		ymax=1
		]
		
		\addplot[color=mittelblau, solid, mark=x, mark options={solid}, line width = 1pt]
		table[col sep=comma]{
		0.170,7.641e-10
		0.180,8.955e-09
		0.190,8.984e-08
		0.200,7.828e-07
		0.210,5.895e-06
		0.220,3.787e-05
		0.230,2.057e-04
		0.240,9.444e-04
		0.250,3.690e-03
		0.260,1.239e-02
		0.270,3.607e-02
		0.280,9.134e-02
		0.290,2.006e-01
		0.300,3.778e-01
			};
		\label{plot:desc}
		\addlegendentry{\footnotesize SC, 5 bit (DE optimized code)};
		
		\addplot[color=rot, dotted, mark=+, mark options={solid}, line width = 1pt]
		table[col sep=comma]{
		0.100,1.047e-09
		0.110,2.524e-08
		0.120,4.009e-07
		0.130,4.711e-06
		0.140,4.194e-05
		0.150,2.794e-04
		0.160,1.400e-03
		0.170,5.428e-03
		0.180,1.687e-02
		0.190,4.353e-02
		0.200,9.606e-02
		0.210,1.854e-01
		0.220,3.174e-01
		0.230,4.860e-01
		0.240,6.676e-01
		0.250,8.261e-01
		0.260,9.318e-01
		0.270,9.818e-01
		0.280,9.970e-01
		0.290,9.997e-01
		0.300,1.000e+00
	};
		\label{plot:sc}
		\addlegendentry{\footnotesize SC, 5 bit};
		
		\addplot[color=rot, solid, mark=o, mark options={solid}, line width = 1pt]
		table[col sep=comma]{
		0.3,	0.2209
		0.29,	0.08955
		0.28,	0.02542
		0.27,	0.005381
		0.26,	0.0007324
		0.25,	0.00007202
		0.24,	0.000005019
		0.23,	0.0000003659
		0.22,	0.00000003521
		0.21,	0.00000000314
		};
		\label{plot:indep}
		\addlegendentry{\footnotesize AE-SC-32 Independent (32 int.)};
		
		\addplot[color=orange, solid, mark=pentagon, mark options={solid}, line width = 1pt]
		table[col sep=comma]{
			0.3,	0.2191
			0.29,	0.08872
			0.28,	0.02581
			0.27,	0.005254
			0.26,	0.0007557
			0.25,	0.00007135
			0.24,	0.000005397
			0.23,	0.0000003771
			0.22,	0.00000003217
			0.21,	0.00000000337
		};
		\label{plot:casc}
		\addlegendentry{\footnotesize AE-SC-32 Cascaded (5 int.)};
		
		\addplot[color=apfelgruen, solid, mark=*, mark options={solid}, line width = 1pt, mark size=1pt]
		table[col sep=comma]{
		0.3,	0.2262
		0.29,	0.08647
		0.28,	0.02695
		0.27,	0.005637
		0.26,	0.0007428
		0.25,	0.00007833
		0.24,	0.000005405
		0.23,	0.0000004042
		0.22,	0.00000003383
		};
		\label{plot:rec}
		\addlegendentry{\footnotesize AE-SC-32 Recursive (1 int.)};

		\addplot[color=apfelgruen, dashed, mark=*, mark options={solid}, line width = 1pt, mark size=1pt]
		table[col sep=comma]{
		0.3,	0.07397
		0.29,	0.0197
		0.28,	0.004081
		0.27,	0.0005092
		0.26,	0.00007288
		0.25,	0.00001061
		0.24,	0.00000168
		};
		\label{plot:rec_127}
		\addlegendentry{\footnotesize AE-SC-127 Recursive (1 int.)};

		\addplot[color=black, densely dashed, line width = 0.5pt]
		table[col sep=comma]{
		0.200,5.219e-10
		0.210,5.407e-09
		0.220,4.799e-08
		0.230,3.695e-07
		0.240,2.497e-06
		0.250,1.494e-05
		0.260,7.991e-05
		0.270,3.847e-04
		0.280,1.678e-03
		0.290,6.676e-03
		0.300,2.434e-02
		};
	\label{plot:ub}
\addlegendentry{\footnotesize Truncated Union Bound};
	
	\coordinate (x1) at (axis cs:0.12,2.1e-7) {};
	\coordinate (x2) at (axis cs:0.226,2.1e-7) {};
	\coordinate (x3) at (axis cs:0.191,1e-7) {};
	\coordinate (x4) at (axis cs:0.223,1e-7) {};
	\end{axis}
	\draw[draw=none, rot]   (x1) -- node[midway,above=1pt,fill=white, fill opacity=0.75, text opacity=1,inner sep=0pt] {\footnotesize AED gain (same code)} (x2);
	\draw[draw=none, mittelblau]   (x3) -- node[midway,below=1pt,align=left,fill=white, fill opacity=0.75, text opacity=1,inner sep=0pt]{\footnotesize opt. SC vs.\\[-5pt]\footnotesize opt. AED} (x4);
	\draw[rot,thick,-Latex] (x1) -- (x2);
	\draw[mittelblau,thick,-Latex] (x3) -- (x4);
\end{tikzpicture}

%% file: fig/interleaver_area.tex
\begin{tikzpicture}
	\begin{axis}[
		ymin=0,
		ymax=14,
		ybar,
		height=.5\columnwidth,
		width=0.95\columnwidth,
		enlarge x limits=0.18,
		ymajorgrids,
		xtick style={draw=none}, %
		nodes near coords style={
			/pgf/number format/fixed,
			/pgf/number format/precision=2
		},
		legend cell align={left},
		legend image code/.code={
			\draw [#1] (0cm,-0.1cm) rectangle (0.2cm,0.25cm); },
		ylabel={Area [fp64 multipliers]},
		nodes near coords,
		symbolic x coords={Independent, Cascaded, Recursive},
		xtick=data]
		\addplot[fill=mittelblau!50, bar shift=-4mm] coordinates {(Independent,11.30) (Cascaded, 1.42) (Recursive, 0.19)};
		\addplot[fill=apfelgruen!50, bar shift=4mm] coordinates {(Independent,5.65) (Cascaded, 0.71) (Recursive, 0.09)};
		\addlegendentry{Full}
		\addlegendentry{Blockwise ($b=2$)}
	\end{axis}
\end{tikzpicture}

%% file: main.bbl
\begin{thebibliography}{10}
\providecommand{\url}[1]{#1}
\csname url@samestyle\endcsname
\providecommand{\newblock}{\relax}
\providecommand{\bibinfo}[2]{#2}
\providecommand{\BIBentrySTDinterwordspacing}{\spaceskip=0pt\relax}
\providecommand{\BIBentryALTinterwordstretchfactor}{4}
\providecommand{\BIBentryALTinterwordspacing}{\spaceskip=\fontdimen2\font plus
\BIBentryALTinterwordstretchfactor\fontdimen3\font minus
  \fontdimen4\font\relax}
\providecommand{\BIBforeignlanguage}[2]{{%
\expandafter\ifx\csname l@#1\endcsname\relax
\typeout{** WARNING: IEEEtran.bst: No hyphenation pattern has been}%
\typeout{** loaded for the language `#1'. Using the pattern for}%
\typeout{** the default language instead.}%
\else
\language=\csname l@#1\endcsname
\fi
#2}}
\providecommand{\BIBdecl}{\relax}
\BIBdecl

\bibitem{suh2007physical}
G.~E. Suh and S.~Devadas, ``{Physical Unclonable Functions for Device
  Authentication and Secret Key Generation},'' in \emph{Proceedings of the 44th
  Annual Design Automation Conference}, 2007, pp. 9--14.

\bibitem{gunlu2019codeconstruction}
O.~Günlü, O.~İşcan, V.~Sidorenko, and G.~Kramer, ``{Code Constructions for
  Physical Unclonable Functions and Biometric Secrecy Systems},'' \emph{IEEE
  Transactions on Information Forensics and Security}, vol.~14, no.~11, pp.
  2848--2858, 2019.

\bibitem{bloch2021overview}
M.~Bloch, O.~Günlü, A.~Yener, F.~Oggier, H.~V. Poor, L.~Sankar, and R.~F.
  Schaefer, ``{An Overview of Information-Theoretic Security and Privacy:
  Metrics, Limits and Applications},'' \emph{IEEE Journal on Selected Areas in
  Information Theory}, vol.~2, no.~1, pp. 5--22, 2021.

\bibitem{delvaux2015helperdataalgos}
J.~Delvaux, D.~Gu, D.~Schellekens, and I.~Verbauwhede, ``{Helper Data
  Algorithms for PUF-Based Key Generation: Overview and Analysis},'' \emph{IEEE
  Transactions on Computer-Aided Design of Integrated Circuits and Systems},
  vol.~34, no.~6, pp. 889--902, 2015.

\bibitem{maes2016biasedpufs}
\BIBentryALTinterwordspacing
R.~Maes, V.~van~der Leest, E.~van~der Sluis, and F.~M.~J. Willems, ``{Secure
  Key Generation from Biased PUFs: Extended Version},'' \emph{Journal of
  Cryptographic Engineering}, vol.~6, no.~2, pp. 121--137, 2016. [Online].
  Available: \url{https://doi.org/10.1007/s13389-016-0125-6}
\BIBentrySTDinterwordspacing

\bibitem{arikan2009}
E.~Arıkan, ``{Channel Polarization: A Method for Constructing
  Capacity-Achieving Codes for Symmetric Binary-Input Memoryless Channels},''
  \emph{IEEE Trans. on Inf. Theory}, vol.~55, no.~7, pp. 3051--3073, 2009.

\bibitem{chen2017robustpolarpuf}
B.~Chen, T.~Ignatenko, F.~M.~J. Willems, R.~Maes, E.~van~der Sluis, and
  G.~Selimis, ``{A Robust SRAM-PUF Key Generation Scheme Based on Polar
  Codes},'' in \emph{2017 IEEE Global Comm. Conf. (GLOBECOM)}, 2017.

\bibitem{geiselhart2021aedrm}
M.~Geiselhart, A.~Elkelesh, M.~Ebada, S.~Cammerer, and S.~ten Brink,
  ``{Automorphism Ensemble Decoding of Reed--Muller Codes},'' \emph{IEEE
  Transactions on Communications}, vol.~69, no.~10, pp. 6424--6438, 2021.

\bibitem{geiselhart2021polaraut}
------, ``{On the Automorphism Group of Polar Codes},'' in \emph{2021 IEEE Int.
  Symp. on Information Theory (ISIT)}, 2021, pp. 1230--1235.

\bibitem{pillet2021polarcodesaed}
C.~Pillet, V.~Bioglio, and I.~Land, ``{Polar Codes for Automorphism Ensemble
  Decoding},'' in \emph{IEEE Inf. Theory Workshop (ITW)}, 2021.

\bibitem{kestel2023URLLC}
C.~Kestel, M.~Geiselhart, L.~Johannsen, S.~ten Brink, and N.~Wehn,
  ``{Automorphism Ensemble Polar Code Decoders for {6G} {URLLC}},'' in
  \emph{Int. ITG Workshop on Smart Antennas (WSA) and Conf. on Systems,
  Commun., and Coding (SCC)}, Braunschweig, Germany, Mar. 2023.

\bibitem{ren2024highthroughputbpl}
Y.~Ren, Y.~Shen, L.~Zhang, A.~T. Kristensen, A.~Balatsoukas-Stimming,
  E.~Boutillon, A.~Burg, and C.~Zhang, ``{High-Throughput and Flexible Belief
  Propagation List Decoder for Polar Codes},'' \emph{IEEE Transactions on
  Signal Processing}, vol.~72, pp. 1158--1174, 2024.

\bibitem{li2025routes}
J.~Li, H.~Zhou, R.~Seah, and W.~Gross, ``{Automorphism Ensemble Decoding of
  Polar Codes with Reduced Number of Routes},'' in \emph{2025 13th Int. Symp.
  on Topics in Coding (ISTC)}.\hskip 1em plus 0.5em minus 0.4em\relax IEEE, Aug
  2025, pp. 1--5.

\bibitem{bardet2016algebraicproperties}
M.~{Bardet}, V.~{Dragoi}, A.~{Otmani}, and J.~{Tillich}, ``{Algebraic
  Properties of Polar Codes From a New Polynomial Formalism},'' in \emph{IEEE
  Inter. Symp. Inf. Theory (ISIT)}, 2016, pp. 230--234.

\bibitem{schnabl1995rmgcc}
G.~Schnabl and M.~Bossert, ``{Soft-decision Decoding of Reed--Muller Codes as
  Generalized Multiple Concatenated Codes},'' \emph{IEEE Transactions on
  Information Theory}, vol.~41, no.~1, pp. 304--308, 1995.

\bibitem{simplifiedSC}
A.~Alamdar-Yazdi and F.~R. Kschischang, ``{A Simplified Successive-Cancellation
  Decoder for Polar Codes},'' \emph{IEEE Communications Letters}, vol.~15,
  no.~12, pp. 1378--1380, 2011.

\bibitem{bioglio2023groupproperties}
V.~Bioglio, I.~Land, and C.~Pillet, ``{Group Properties of Polar Codes for
  Automorphism Ensemble Decoding},'' \emph{IEEE Transactions on Information
  Theory}, vol.~69, no.~6, pp. 3731--3747, 2023.

\bibitem{serre1977representations}
J.-P. Serre, \emph{Linear Representations of Finite Groups}.\hskip 1em plus
  0.5em minus 0.4em\relax New York: Springer-Verlag, 1977, translated from the
  second French edition by Leonard L. Scott, Graduate Texts in Mathematics,
  Vol. 42.

\bibitem{hestenes1970singer}
M.~D. Hestenes, ``{Singer Groups},'' \emph{Canadian Journal of Mathematics},
  vol.~22, no.~3, pp. 492--513, 1970.

\bibitem{pillet2022classification}
C.~Pillet, V.~Bioglio, and I.~Land, ``{Classification of Automorphisms for the
  Decoding of Polar Codes},'' in \emph{ICC 2022 - IEEE International Conference
  on Communications}, 2022, pp. 110--115.

\bibitem{ye2022completeinvariant}
Z.~Ye, Y.~Li, H.~Zhang, R.~Li, J.~Wang, G.~Yan, and Z.~Ma, ``{The Complete
  SC-Invariant Affine Automorphisms of Polar Codes},'' in \emph{2022 IEEE Int.
  Symp. on Information Theory (ISIT)}, 2022, pp. 2368--2373.

\bibitem{mori2009DE}
R.~Mori and T.~Tanaka, ``{Performance and Construction of Polar Codes on
  Symmetric Binary-Input Memoryless Channels},'' in \emph{2009 IEEE
  International Symposium on Information Theory}, 2009, pp. 1496--1500.

\bibitem{geiselhart2023ratecompatible}
M.~Geiselhart, J.~Clausius, and S.~ten Brink, ``{Rate-Compatible Polar Codes
  for Automorphism Ensemble Decoding},'' in \emph{2023 12th International
  Symposium on Topics in Coding (ISTC)}, 2023, pp. 1--5.

\end{thebibliography}
